\documentclass[12pt,a4paper]{iopart}

\usepackage{rotating,graphicx}
\usepackage{amssymb}
\usepackage{amsfonts}%
\usepackage{caption}
\usepackage[utf8x]{inputenc}
\usepackage{lmodern,textcomp}
\usepackage[english]{babel} 
\usepackage[T1]{fontenc}
\usepackage{tabularx} 
\usepackage{color}
\usepackage[colorlinks]{hyperref}
\usepackage{setspace}
\usepackage{layouts}
\begin{document}

\title[A feasibility study for a DR system in JT-60SA]{A feasibility study for a Doppler Reflectometer System in the JT-60SA tokamak}
\author{D. Carralero$^1$, T. Happel$^2$, T. Estrada$^1$, T. Tokuzawa$^{3,4}$, J. Martínez$^1$, E. de la Luna$^1$, A. Cappa$^1$ and J. García $^5$}
\address{$^1$ Laboratorio Nacional de Fusión. CIEMAT, 28040 Madrid, Spain.}
\address{$^2$ Max-Planck-Institut für Plasmaphysik, D-85748, Garching, Germany.}
\address{$^3$National Institute for Fusion Science, 322-6 Oroshi-cho, Toki 509-5292, Japan} 
\address{$^4$SOKENDAI, 322-6 Oroshi-cho, Toki 509-5292, Japan} 
\address{$^5$CEA, Institute for Magnetic Fusion Research, 13108 Saint-Paul-Lez-Durance, France} 

\ead{daniel.carralero@ciemat.es}

\begin{abstract}

In this work we present a study on the viability and practicality of installing a Doppler reflectometer (DR) system in the JT-60SA advanced tokamak. First, we discuss its scientific scope in the context of the JT-60SA research plan. We identify a number of fields in which a DR would be very relevant for the accomplishment of said plan and outline a scientific program for the diagnostic. Then, starting from a number of design hypothesis, we use a ray tracing code to carry out a feasibility study for a number of relevant scenarios and identify a geometric solution for the installation of a DR such that both core and edge can be probed in the prescribed wave number range, thus achieving the proposed scientific objectives. Finally, we perform a preliminary discussion on the different possibilities for a conceptual design  (including a minimum viable system and a baseline system) and their requirements in terms of components and space. We conclude that a viable conceptual design could be carried out using a small fraction of a horizontal port, leaving room for additional diagnostic systems. 

\end{abstract}

\maketitle

\section{Introduction}

The JT-60SA tokamak has been designed as part of the Broader Approach agreement to fusion power, with the main objective of reproducing relevant scenarios from which the operation of ITER can be anticipated and the demonstration of integrated performance scenarios for ITER and DEMO. In particular, the JT-60SA Research Plan \cite{ref2} states that:
\begin{quotation}
	"The validation of theoretical models and simulation codes with the aim of establishing a solid basis for the design of ITER and DEMO scenarios is one of the main objectives of the JT-60SA scientific program."
\end{quotation}

One of the major uncertainties when predicting future ITER and DEMO scenarios is anomalous heat, particle and momentum transport \cite{ref3}, dominated by turbulent processes which will take place under plasma parameters ($\beta$,$\nu^*$, $\rho^*$, etc.) that can´t be reached in current day machines. Since first-principles turbulence simulations based on gyrokinetic (GK) theory \textcolor{black}{-using codes such as EUTERPE \cite{Jost01, Kornilov05}, GENE\cite{Jenko00}, GS2\cite{Dorland00} or stella \cite{Barnes19}-} are extremely time consuming, this anomalous transport has been so far introduced in integrated modeling codes with transport coefficients obtained from reduced transport models like GLF23 \cite{ref4} or CDBM \cite{ref5}. However, these codes have been validated using only data from experiments in JT-60U and JET \cite{ref6,ref7} which are still far from some of the relevant regimes. In this context, the lack of a sufficiently validated fast model of turbulence makes it difficult to confidently predict phenomena like the formation of a transport barrier in the core (ITB) or the impact of fast ions on turbulence \cite{ref8}, thus making the comparison of detailed experimental observations of the characteristics of the turbulence with GK codes one of the central modelling research needs for ITER and DEMO \cite{ref2}.\\

The importance of validating turbulence models in conditions relevant to ITER and DEMO highlights the relevance of JT-60SA as the perfect benchmark for this kind of studies: Indeed, this tokamak will achieve plasmas with features much closer to ITER and DEMO than those in present-day machines, such as realistic heating conditions (dominant electron heating, low central fueling and low external torque), a significant fast ion fraction, high beta vales or highly shaped plasmas. This evaluation of fundamental turbulent transport mechanisms is relevant for a wide number of cases from the scientific program of JT-60SA, ranging from the aforementioned ITB formation to the general performance of the machine under different scenarios and the L-H confinement mode transition and the associated formation of the edge transport barrier (ETB) at the pedestal. The obvious consequence of this is that, in order to carry out such validation of turbulent models in JT-60SA, suitable diagnostics will be required for a systematic characterization of turbulence. In particular, it will be required that this characterization is detailed enough for meaningful comparison with simulations from GK models. One of the best diagnostics available for this mission, and one which has already been proposed for its installation in JT-60SA \cite{ref2,ref9}, is a Doppler reflectometer.\\ 

The present work describes a feasibility study for the installation of a Doppler Reflectometer (DR) system in the JT-60SA tokamak. In Section \ref{Sec1}, we first discuss the relevance of such diagnostic in the experimental program of JT-60SA and outline the main elements of its possible scientific mission. After this, in Section \ref{Sec2} we present the core of the feasibility study, in which ray tracing code TRAVIS \cite{ref32} is used to determine an optimal geometry for the DR, which would allow it to conduct the proposed observations under the main scenarios foreseen in JT-60SA operation. Taking this result, a preliminary conceptual design is proposed in Section \ref{Sec3}, including a discussion of the different options for its development depending on port space available for the diagnostic. Finally, the main conclusions of the study are outlined in Section \ref{Sec4}. This feasibility study is part of a more general proposal \cite{ref1}, involving the design and implementation of a DR in JT-60SA \textcolor{black}{as well as the  proposal of a scientific program}, but also covering its operation, data analysis, GK modeling, etc.. This proposal was drafted along several scientific teams with well-established expertise in each of those stages of the diagnostic scientific exploitation.

\section{Scientific Mission}\label{Sec1}

A Doppler Reflectometer \textcolor{black}{emits a microwave beam into the plasma which, unlike the one found in conventional reflectometers, is launched with an oblique angle with respect to its cut-off surface. Hence, the diagnostic measures the scattering of the inbound beam caused by density fluctuations at the cut-off layer with a perpendicular wavenumber, $k_\perp$, defined by the Bragg criterion, $k_\perp =  2k_0\sin{\alpha_0}$, where $k_0$ and $\alpha_0$ are respectively the microwave beam wavenumber and incidence angle (with $\alpha_0 = 0$ meaning perpendicular incidence with respect to the local flux surface) \cite{Hirsch01}. The power of the scattered signal, $P_{scatt}$ is related to the amplitude of the density fluctuations of wavenumber k$_\perp$, $\delta n(k_\perp)$. In particular, under certain approximations (such as low amplitude turbulence), it can be approximated as $P_{scatt} \propto \delta n(k_\perp)^2$ \cite{Gusakov04,Blanco08}. The frequency of this scattered beam is Doppler-shifted due to the velocity of the associated fluctuations. Usualy, radial and parallel components can be neglected if the incident beam is correctly aligned, so the measurement of this Doppler shift, $\omega_D$, can be used to estimate the perpendicular velocity of turbulence $\omega_D = u_\perp k_{\perp}$. Nevertheless, it can be shown that a proper alignment is critical for DR measurements, as the scattered power decays rapidly with the mismatch angle, $\gamma_{mis}$, defined as ${\bf k_i}\cdot {\bf B} = \cos{\pi/2-\gamma_{mis}}$, where $\bf{k_i}$ is the wave vector of the incident beam at the cut-off layer \cite{Hillesheim15}. A practical criterion to achieve small values of $\gamma_{mis}$ which has been used in the literature is to impose $|k_\parallel/k_\perp| < 0.1$ at the cut-off layer \cite{Conway11}. 
	 Given the described characteristics, a DR} presents a number of features which make it rather adequate for the characterization of turbulence:
\begin{itemize}
\item [-] Valid for quantitative analysis: k$_\perp$-selective, radial scans of the amplitude of local density fluctuations. By correlation studies, average size and tilt of turbulent structures can be measured \cite{ref10}.
\item [-] Wave number spectra measurements: by using a steerable mirror \textcolor{black}{(allowing the selection of the $\alpha_0$ value)}, a range of k$_\perp$ values can be probed. Typical values of interest are in the \textcolor{black}{ k$_\perp\rho_i \simeq$ 0.3-5 range}, appropriated to observe ITG and TEM turbulence.
\item [-] \textcolor{black}{Adequate spatial, spectral and  temporal resolution: typical values for the first two are $\Delta$r $\simeq$ 1 cm and $\Delta$ k$_\perp \simeq$ 1 cm$^{-1}$. Regarding the last, u$_\perp$ measurements require sufficient data for a proper spectral analysis, typically resulting in an effective time resolution below $1$ ms (although u$_\perp$ oscillations can be measured at substantially shorter time scales, up to the hundreds of kHz \cite{Estrada12}). Instead, fluctuation amplitude measurements are in principle only limited by the data sampling (usually above the MHz resolution, sufficient for the study of turbulent phenomena). }Typical radial range covers SOL, edge and core (specific $\rho$ values for JT-60SA scenarios are discussed later).
\item [-] Well suited for comparison to GK models \cite{Casati09} - \cite{ref12}, including, by implementing a GK model output in a full-wave code, a direct comparison between experimental results and synthetic diagnostics in the simulation \cite{ref13}. 
\end{itemize}

However, the advantages of a DR system are not only limited to its capabilities as a diagnostic for the measurement of fluctuations: it can be used as well for the measurement of radial profiles of perpendicular plasma flows, u$_{\perp}$, which are relevant for one of the central research needs for ITER and DEMO (and therefore one of the main objectives of JT-60SA), such as the validation of current models of Neoclassical Toroidal Viscosity (NTV) at high $\beta$, high intrinsic rotation conditions \cite{ref2}. In high $\beta$ plasmas with DEMO-equivalent highly shaped configuration, toroidal rotation (and rotation shear) is expected to play a critical role as a stabilization mechanism in JT-60SA and even a small breaking of axisymmetry can produce non-negligible toroidal viscosity and significant reduction in the achievable toroidal rotation levels. This effect is already non-negligible in current tokamaks (e.g., in AUG \cite{ref14} or JT-60U \cite{ref15,ref16}), but it may represent a much more serious challenge in highly self-regulated ones in which the external torque represents only a small fraction of the rotation balance. In this context, by obtaining u$_{toroidal}$  from u$_{\perp}$ (neoclassical flow contributions can be calculated in a tokamak), a DR system can be used to carry out studies in rotation and neoclassical viscosity as a diagnostic complementary to CXRS, with the advantage of being independent of NBI heating levels and penetration.\\ 

Finally, measurements of u$_\perp$ can be used to estimate E$_r$ profiles: \textcolor{black}{If simulations are used to estimate the phase velocity of the $k_\perp$- selected fluctuations, u$_{phase}$, or u$_{phase} \ll u_{E\times B}$ is simply assumed} (as it is typically the case in most experimental settings), it is possible \textcolor{black}{to measure simultaneously E$_r$ and E$_r$ shear profiles along with those of fluctuation amplitudes}. These measurements are of great importance for the aforementioned validation of transport models, as one of the key open questions in them is the stabilizing effect of radial electric fields in turbulence, generally associated to the decorrelation of turbulent structures by the E$_r$ shear. One prominent example of this effect, of great importance for ITER and DEMO, is the role of an E$\times$B velocity shear in the formation of the edge transport barrier (ETB) at the pedestal associated to the L-H confinement mode transition. According to the ITER research plan \cite{ref17}, both the fundamental mechanisms behind the turbulence stabilization associated to H-mode and their parametric dependences must be characterized. In consequence, a DR would be a very useful diagnostic in the significant fraction of experimental time in JT-60SA which will be dedicated to study L-H transition parameters under a variety of isotopes, densities, collisionalities and external torque \cite{ref2}.\\

Considering all of the above, we will outline next the three main lines of research on which the scientific mission of a DR system installed in JT-60SA would be centered.\\

\subsection{Turbulence-GK simulation comparison }

A DR can be used to carry out measurements of core turbulence characteristics, namely the amplitude of density fluctuations associated to a given k$_\perp$ determined by the incidence angle and wavelength of the probing beam. If the system has steering capability (see next section), multiple wave numbers in that range can be measured, allowing for the evaluation of the k$_\perp$-spectra of the turbulence \cite{Casati09, ref18, Vermare11, ref19, ref20}. By this means, fluctuation amplitudes and wave number and frequency spectra could be characterized and compared with both analytical models \cite{ref21} and gyro-kinetic simulations in which appropriate synthetic diagnostics have been implemented. The possible comparisons range from linear stability analysis to identify the general type of turbulence to the more complex and computationally demanding prediction of the fluctuation spectrum and amplitude by non-linear simulations. DR data has already been used to carry out this kind of comparisons in tokamaks, such as the studies carried out in AUG \cite{ref13}, where the turbulence field has been calculated using flux-tube, non-linear GENE simulations. Using said field as an input for two-dimensional full-wave code, a synthetic DR output is obtained, thus allowing for a direct comparison of DR output and GK turbulence predictions. Once GK codes have been validated by comparison with experimental data, energy, momentum and particle transport values can be obtained from these codes, either directly or in the form of parameter-dependent transport coefficients, and compared to predictions from integrated modelling codes. With this approach, turbulent transport could be systematically characterized for the accessible range of non-dimensional parameters $\rho^*$, $\nu^*$ and $\beta$, and the limits of current predictive capabilities could be assessed and eventually overcome.

\subsection{Edge turbulence and L-H transition}

The complex interaction between turbulence, zonal flows (ZF) and mean E$_r$ has been an active area of study by Doppler reflectometry in recent years: Although first observations come from the stellarator community \cite{ref22,ref23}, this diagnostic has been central in the observation of predator-prey dynamics before the L-H transition in tokamaks \cite{ref24}, and has been used to characterize the minimum values of the E$_r$ well \cite{Schirmer06, ref25} and to detect stationary ZFs \cite{ref26}. This kind of studies can be extended to the first phases of operation of JT-60SA, which will address the characterization of the H-mode access conditions: measurements of E$_r$ and turbulence amplitude across the confinement transition could be carried out during the characterization of the L-H power threshold for a number of different densities, magnetic field values, etc.. During this first analysis, the presence of limit-cycle oscillations and predator-prey interactions between flows and turbulence could be analysed. As well, fine radial E$_r$ scans could be conducted in order to search for stationary ZF such as the ones detected in JET. A second DR channel installed in the same line (see next section) would allow the simultaneous measurement of E$_r$ at two radial positions, thus allowing the observation of the evolution of the E$_r$ shear and its influence in the ETB formation \cite{ref27,Schirmer07}. Also, a second channel installed in a different toroidal and/or poloidal position could be used to detect non-stationary ZFs and measure their evolution across the confinement transition \cite{Hillesheim12}.  Besides, the minimum value of E$_r$ at the bottom of the well can be measured and the existence of an E$_r$/B threshold could be determined by combining data from different operational regimes. After this characterization has been carried out, E$_r$, E$_r$ shear and zonal flow values can be compared to kinetic predictions in order to assess the likelihood of a turbulent origin of the ETB formation. Eventually, measurements of ZFs can used to validate the predictive capabilities of GK codes such as EUTERPE, GENE, ORB5, etc. 
Finally, another important topic for JT-60SA is the understanding and control development of small/no ELM operation modes such as grassy ELM or QH-mode \cite{ref2, Viezzer18,Labit19}. In this context, the DR can be used for the characterization of fluctuations in the pedestal during the parametric scans of these regimes in order to investigate the role of turbulence and E$_r$ in their operational window.

\subsection{Rotation and NTV}

Measurements provided by the DR system would be of great importance in a device dedicated to rotation studies, such as JT-60SA. The proposed system would allow cross-validation with the CXRS diagnostic across most of the radial profile, which can be particularly important under high density conditions, in which the temporal resolution and accuracy of the second might be reduced due to lower beam penetration and Bremsstrahlung issues. In the first place, measurements of rotation in the edge can be carried out during experiments making use of ripple inducing control mechanisms (RMPs, RWMCs, etc.) in order to carry out a detailed characterization of the interplay between different momentum injection and symmetry-breaking mechanisms, including turbulent ones \cite{ref28}. Once the module of the torque associated to the neoclassical toroidal viscosity (NTV) has been identified for different perturbations to the magnetic configuration and plasma profiles, neoclassical codes FORTEC-3D or KNOSOS \cite{ref29} could be used to estimate the values expected according to neoclassical theory, and compare them to measurements. FORTEC-3D simulations of the NTV effect in JT-60SA \cite{ref15} predict a strongly sheared edge radial profile of E$_r$ which should appear prominently in DR measurements, allowing for a detailed characterization, which should be repeated under a wide range of relevant parameters (collisionality, E$_r$ profiles, different levels of symmetry breaking, etc.).
In parallel, measurement of the radial electric field may become a key quantity for the characterization of the neoclassical impurity transport if the loss of axisymmetry leads to relevant changes in it. In this sense, beyond the direct measurement of changes in the E$_r$ component, poloidally separated reflectometers and/or steerable mirrors could be used to measure poloidal asymmetries in the plasma potential. This kind of measurements have already been successfully carried out at the stellarator TJ-II \cite{ref30} and have been related to strong corrections in impurity transport predictions \cite{ref31}.

\section{Feasibility Study }\label{Sec2}

The main objective of the feasibility study is to determine the existence of a geometric solution for a DR in JT-60SA. By this, we mean that a set of beams can be found for a given antenna position such that E$_r$ and turbulence amplitude can be measured in the region and range of k$_\perp$ values of interest for the most likely scenarios foreseen for JT-60SA. \textcolor{black}{These beams are defined by two launching angles, which may include the previously defined $\alpha$ and  $\gamma_{mis}$ angles. However, to ease the description of the study, azimuth $\phi$ and elevation $\theta$ with respect to the antenna position will be used instead in the following. In order to carry out these calculations}, we will use the ray tracing code TRAVIS. \\

\subsection{Starting hypothesis}

\begin{figure}
	\centering
	\includegraphics[width=0.85\linewidth] {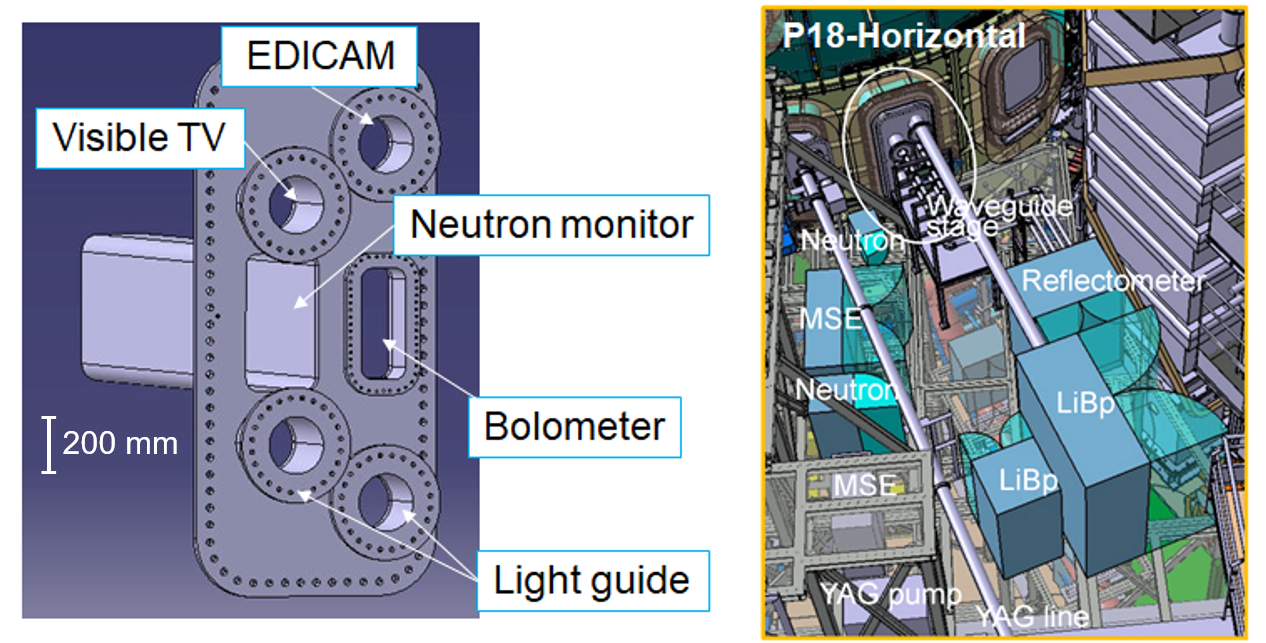}
	\caption{\textit{Overview of the P-18 Horizontal Port \cite{ref32d}. Left, current status of the diagnostic flange, including the diagnostic port allocation. Right, 3D view of the port including an hypothetical reflectometer cabinet and waveguide system.}}
	\label{Figure_2}
\end{figure}

\begin{figure}
	\centering
	\includegraphics[width=0.5\linewidth] {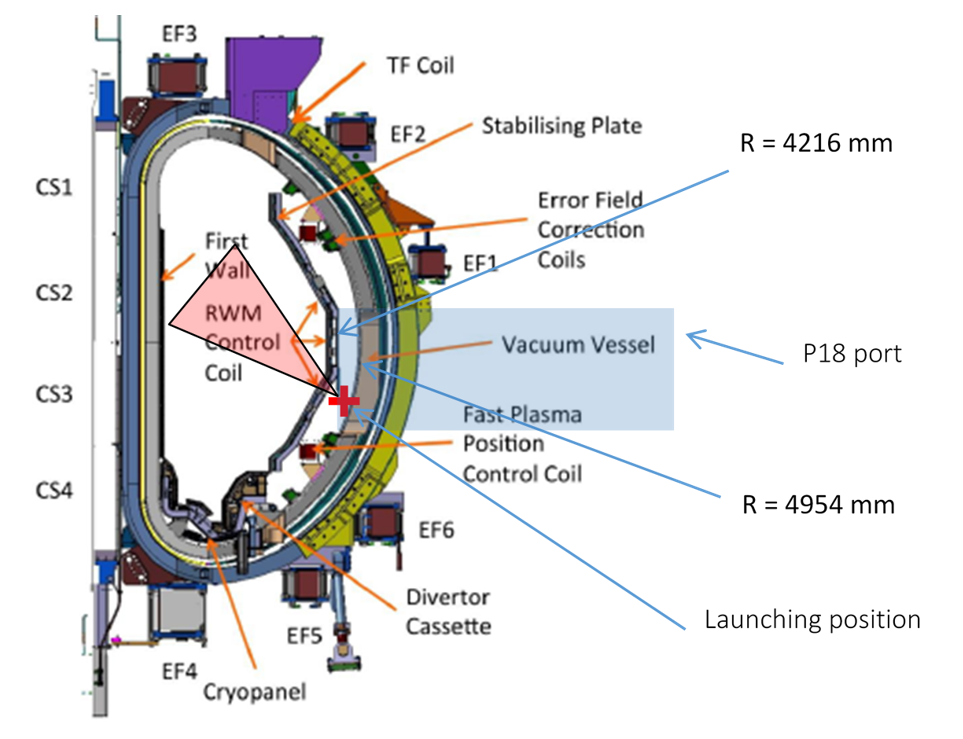}
	\caption{\textit{ \textcolor{black}{Antenna position and launching angles represented on a poloidal section of JT-60SA along with the silhouette of the P18 port (dimensions are approximate).} Image adapted from  \cite{ref34}).}}
	\label{Figure_3}
\end{figure}

As a starting point, we will make a number of hypothesis regarding the launching positions. In particular, following the provisions in the Research Plan \cite{ref2} and the results of several recent discussions \cite{ref32b,ref32c}, we will assume for all effects of the study a DR system installed in the P18-Horizontal diagnostic flange. This flange, displayed in Fig. \ref{Figure_2}, has a number of ports, which have been currently allocated for different diagnostics. We will take the lower ports of the flange, currently dedicated to light guides, as a likely position for the DR system (the specifics of this will be further discussed in the next section, regarding the preliminary conceptual design). Based on this hypothesis, a launching position is determined as displayed in Fig. \ref{Figure_3}: first, the vertical coordinate of the left lower port is taken, and an antenna position is defined between the vacuum vessel and the stabilizing plate, which have respective radial positions of R = 4.216 and 4.954 m. Therefore, the considered coordinates of the antenna are R = 4.5 m and z = -0.64 m. These values are of course tentative, even if the port P18-Horizontal is finally designated for the DR.\\

\subsection{Considered Scenarios}

In order to evaluate the propagation of a microwave beam in the plasma using a ray tracing code, some scenario must be defined regarding a) the magnetic configuration and b) the density and magnetic field profiles (this last aspect is only relevant in the case of X-mode polarization). In order to define these scenarios, we have resorted to the JT-60SA Research Plan \cite{ref2}, from which we have selected the three most representative ones:
\begin{itemize}
\item [-] Full current inductive single null, high density scenario, featuring high Greenwald fraction and plasma current (in the following, high density scenario).
\item [-] Advanced inductive hybrid scenario, featuring moderate densities and plasma currents with high q$_{95}$ values (in the following, hybrid scenario).
\item [-] High-$\beta_N$ full current drive scenario, featuring reduced toroidal field and high Greenwald fraction and q$_{95}$ values (in the following, high-$\beta$ scenario).
\end{itemize} 

\begin{figure}
	\centering
	\includegraphics[width=\linewidth] {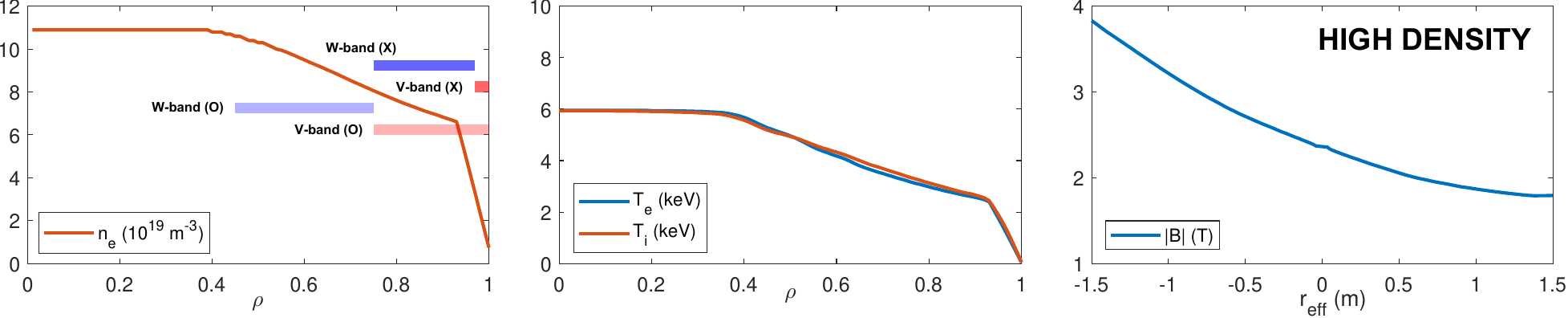}
	\includegraphics[width=\linewidth] {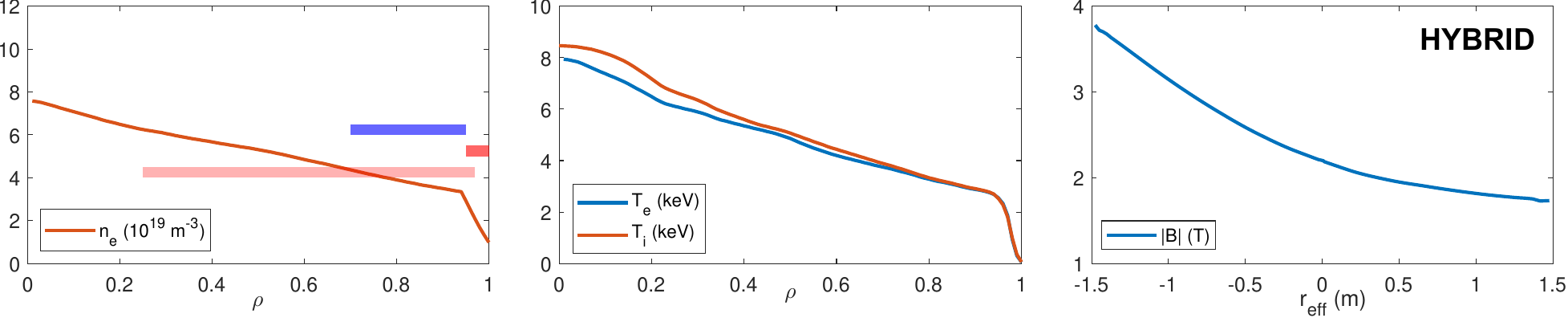}
	\includegraphics[width=\linewidth] {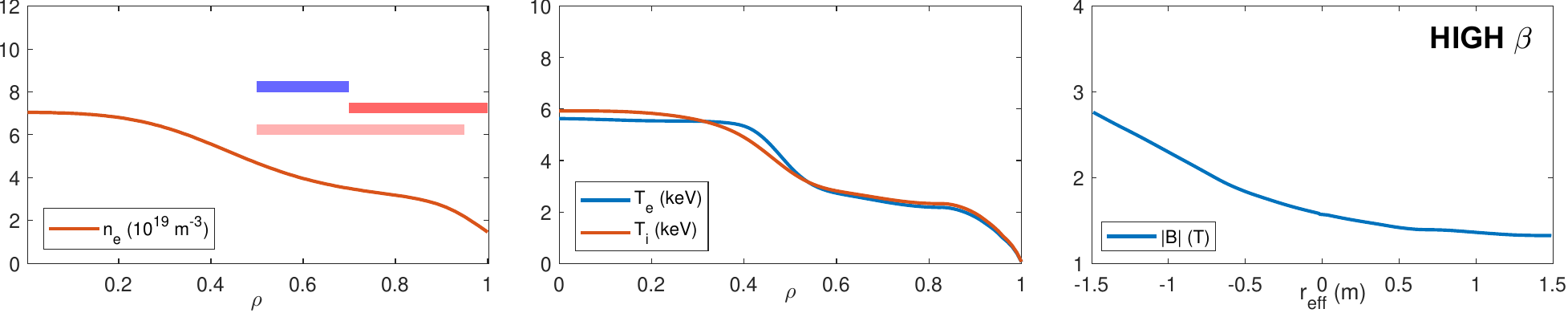}
	\caption{\textit{Plasma and magnetic field profiles of the three considered scenarios. \textcolor{black}{DR accessible regions for k$_\perp \simeq 10$ cm$^{-1}$, as calculated in section \ref{secondTravis}, are represented on the left column using red/blue for V/W band and light/dark shade for O/X mode.}}}
	\label{Figure_4}
\end{figure}

\begin{figure}
\centering
\includegraphics[width=.5\linewidth] {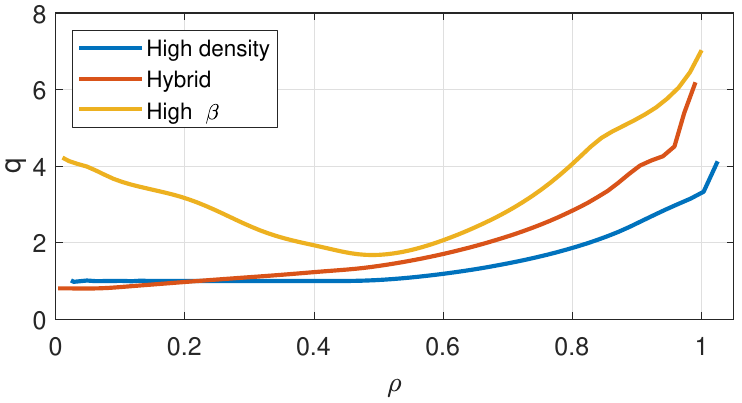}
\caption{\textit{Safety factor profile for each of the three configurations presented in Fig. \ref{Figure_4}.}}
\label{Figure_5}
\end{figure}

 The density, temperature and magnetic field profiles of each of these scenarios are displayed at Fig. \ref{Figure_4}. \textcolor{black}{All of them correspond to plasmas with dominant NBI heating (with positive and negative sources)} and in all cases, the magnetic configuration represents the case in which plasma current and magnetic field at the axis have the same direction. As can be seen, the magnetic field profile of the two first configurations is rather similar, but is substantially reduced in the third. \textcolor{black}{The radial regions accessible to the DR at the k$_\perp \simeq 10$ cm$^{-1}$, which will be calculated in section \ref{secondTravis}, have been represented on the left column. It must be taken into account that, while these results may be considered representative, the whole picture is more complex and these values depend on the particular k$_\perp$ scale being considered, as shown in Figs. \ref{Figure_11} - \ref{Figure_13}.} As well, as can be seen in Fig. \ref{Figure_5}, the safety factor of each of the three configurations is different along the radial region in which the DR is expected to operate, namely  $\rho$ = 0.4 – 1. As a consequence, the optimal angle of incidence to minimize the k$_\parallel$/k$_\perp$ ratio is not the same for all of them, and a compromise must be found in order to conduct angle scans using a single-axis steering mirror. \\

\subsection{First TRAVIS scan: elevation and azimuth angles}

Once the antenna position and the scenarios are defined, ray tracing code TRAVIS has been used to determine optimal launching angles, frequency bands and polarization modes. In order to do this, a scan in elevation and azimuth of the launching angle has been carried out such that the following conditions are met:
\begin{itemize}
\item [-] Minimize the parallel component of the wave number at the cut-off point, k$_{\parallel}$, with respect to the perpendicular to at most, |k$_\parallel$/k$_\perp$| < 0.1. \textcolor{black}{As already discussed, }this is a common design condition for a DR, required to achieve an acceptable S/N ratio in the measurements \cite{Hillesheim15,Conway11}.
\item [-] The signal level of a DR is proportional to the amplitude of the fluctuations which backscatter the beam. Therefore, given the typical ballooning character of turbulence, the backscattering layer should be close to the outer midplane (OMP).
\item [-] The incidence angle should be chosen in such a way that the ITG and TEM turbulence-scales are probed. This means that the k$_\perp\rho_i$ product should be in the \textcolor{black}{0.3-5 range}, where $\rho_i$ is the ion gyroradius at the backscattering layer. 
\end{itemize}

\begin{figure}
	\centering
	\includegraphics[width=.5\linewidth] {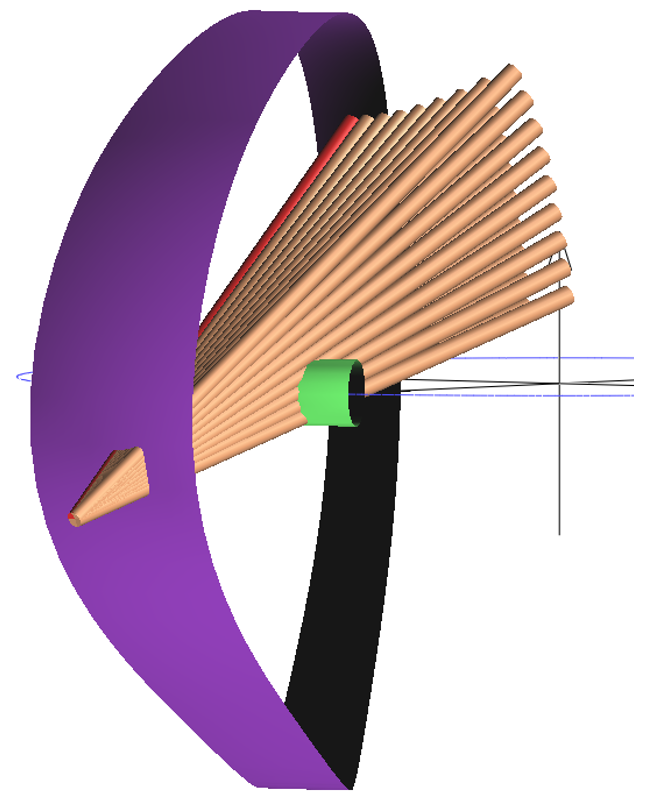}
	\caption{\textit{First scan of elevation and azimuth carried out with TRAVIS. Launching directions of the beam array are represented as golden cones emanating from the antenna position. For reference, purple and green surfaces represent a toroidal section of the $\rho = 1$ and $\rho = 0.1$ flux surfaces, respectively}}
	\label{Figure_6}
\end{figure}
\begin{figure}
	\centering
	\includegraphics[width=\linewidth] {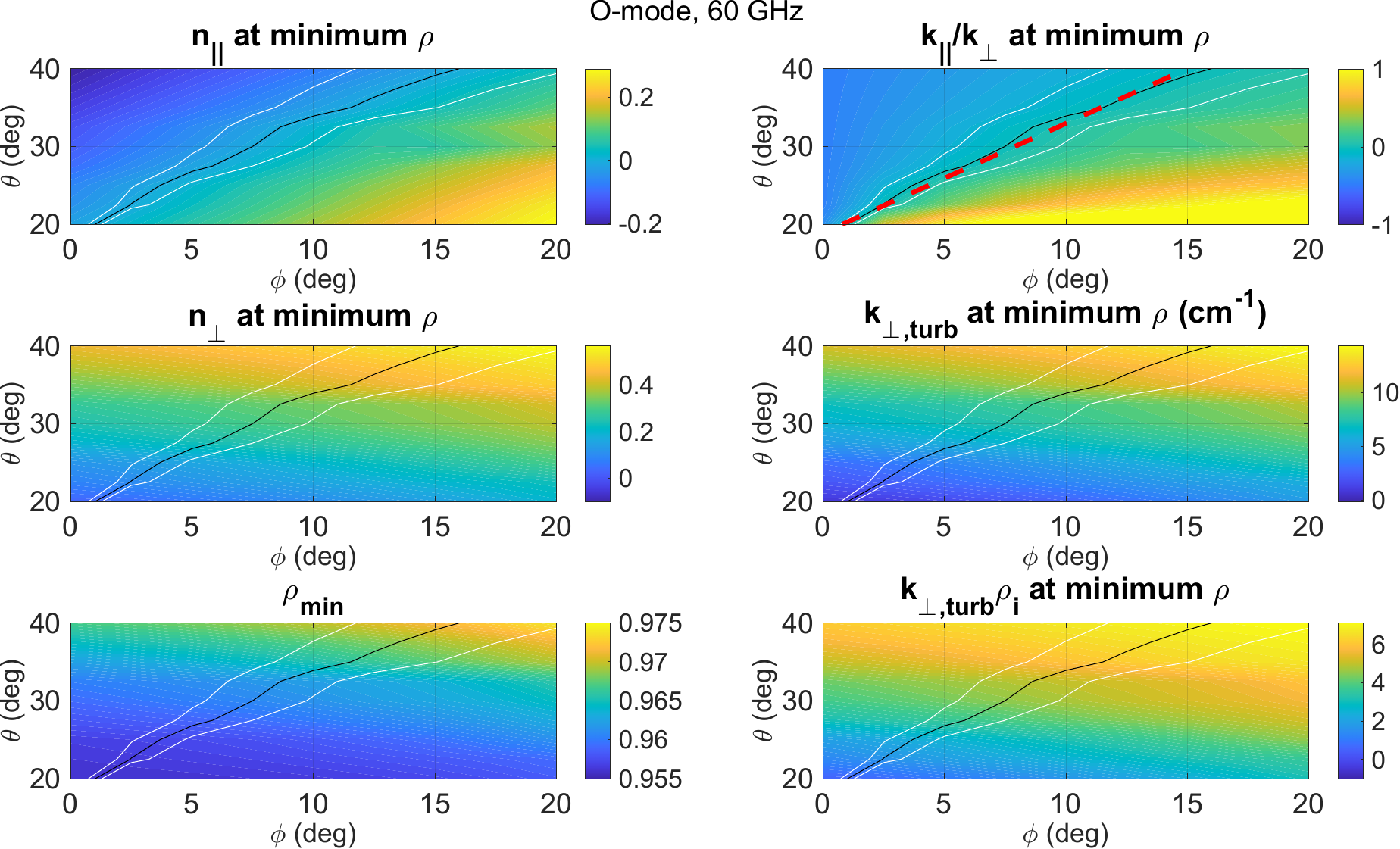}
	\caption{\textit{Scan results for O-mode, 60 GHz beam array. In all figures, $\theta$ and $\phi$ stand for elevation and azimuth, respectively. On the left, from top to bottom parallel, perpendicular refraction indexes and normalized radial position at the cut-off layer. On the right, from top to bottom parallel to perpendicular wave number ratio, absolute value of the perpendicular wavenumber of the turbulence backscattering the beam and k$_\perp \rho_i $ local value at the cut-off layer. Black and white lines correspond to the $\theta$ and $\phi$ values for which k$_\parallel$/k$_\bot$ = 0 and |k$_\parallel$/k$_\bot$| = 0.1, respectively. Red line stands for the $\theta = 1.41 \phi +20$ condition describing the optimal single-axis steerable mirror for this configuration.}}
	\label{Figure_7}
\end{figure}

\begin{figure}
	\centering
	\includegraphics[width=\linewidth] {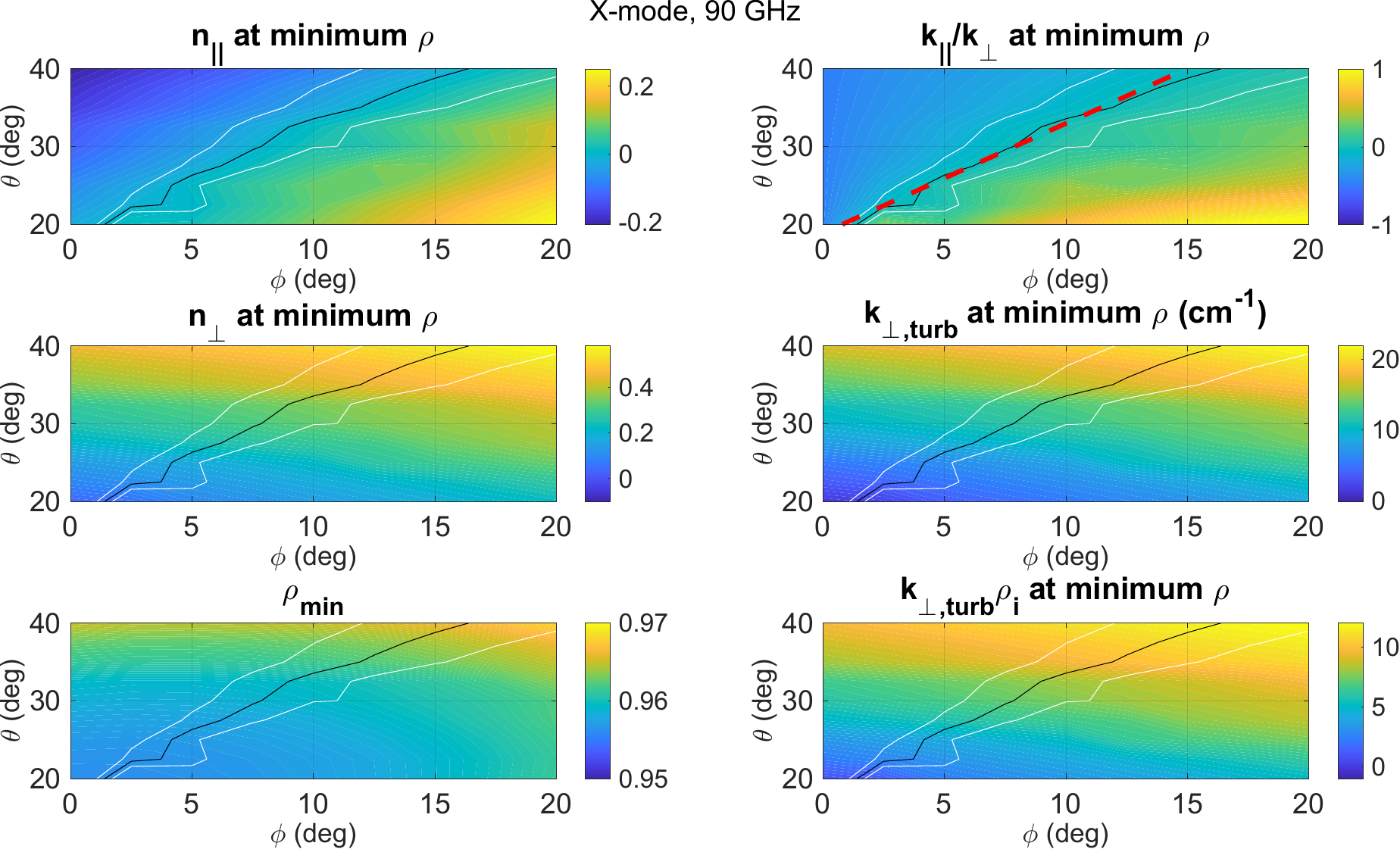}
	\caption{\textit{Scan results for X-mode, 90 GHz beam array. In all figures, $\theta$ and $\phi$ stand for elevation and azimuth, respectively. Plot layout as in Fig. \ref{Figure_7}}}
	\label{Figure_8}
\end{figure}

Taking these constrains into account, a first scan has been carried out in which a matrix of 9x9 rays was launched from the antenna as displayed in Fig. \ref{Figure_6} with elevation angles, $\theta$, ranging in the 20°-40° interval and azimuth angles, $\phi$, ranging in the 0°-20° values. Since the antenna is placed below the equator, a positive elevation is required in order to reach the OMP region. As well, the azimuth angles around 10° provide the best |k$_\parallel$/k$_\perp$| ratios. This scan has been repeated for each scenario, taking a number of frequencies ranging in the V and W frequency bands, as well as for the O and X polarization modes in order to evaluate |k$_\parallel$/k$_\perp$| and k$_\perp\rho_i$ at the different radial positions of the scattering layer. In Figs. \ref{Figure_7} and \ref{Figure_8} two examples of these are presented, in which an array of rays are launched covering the described $\theta$ and $\phi$ range taking O-mode, 60 GHz and X-mode 90  GHz beams in the high density scenario. These two cases can be considered representative of the results of the analysis and allow us to reach the following conclusions:

\begin{itemize}
\item [-] 	In the edge, the radial position of the cut-off layer is mostly determined by the selected polarization and frequency, and only to a lesser extent by the choice of launching angles.
\item [-] 	A reasonable wide corridor of $\theta$ and $\phi$ values exists such that |k$_\parallel$/k$_\perp$| < 0.1 holds in it. \textcolor{black}{This corridor becomes narrower for lower values of $\theta$ and $\phi$}.
\item [-] The k$_\perp$ of the backscattering turbulence depends mostly on the elevation angle. By changing it within the selected range, k$_{\perp,turb}$ can be scanned for around an order of magnitude, taking values in the 1-10 cm$^{-1}$ range.
\item [-]	In the considered scenarios, this means that the probe turbulence is roughly in the k$_\perp\rho_i$ = 0.5-15 range. 
\end{itemize}

From these conclusions, another consequence can be derived: since the |k$_\parallel$/k$_\perp$| < 0.1 result is obtained for launching angles with a roughly constant $\theta$/$\phi$ ratio, the beams yielding optimal measurement cut-off positions are approximately distributed along a plane, which is roughly perpendicular to the magnetic field at the LCFS. This means that a steerable mirror system can be designed capable of probing the whole k$_{\bot,turb}$ range using a single-axis rotation. Such optimal probing space is represented for the high density scenario (in which it can be expressed as $\theta = 1.41 \phi +20$) as a red dashed line in Figs. \ref{Figure_7} and \ref{Figure_8}.\\

\begin{figure}
	\centering
	\includegraphics[width=\linewidth] {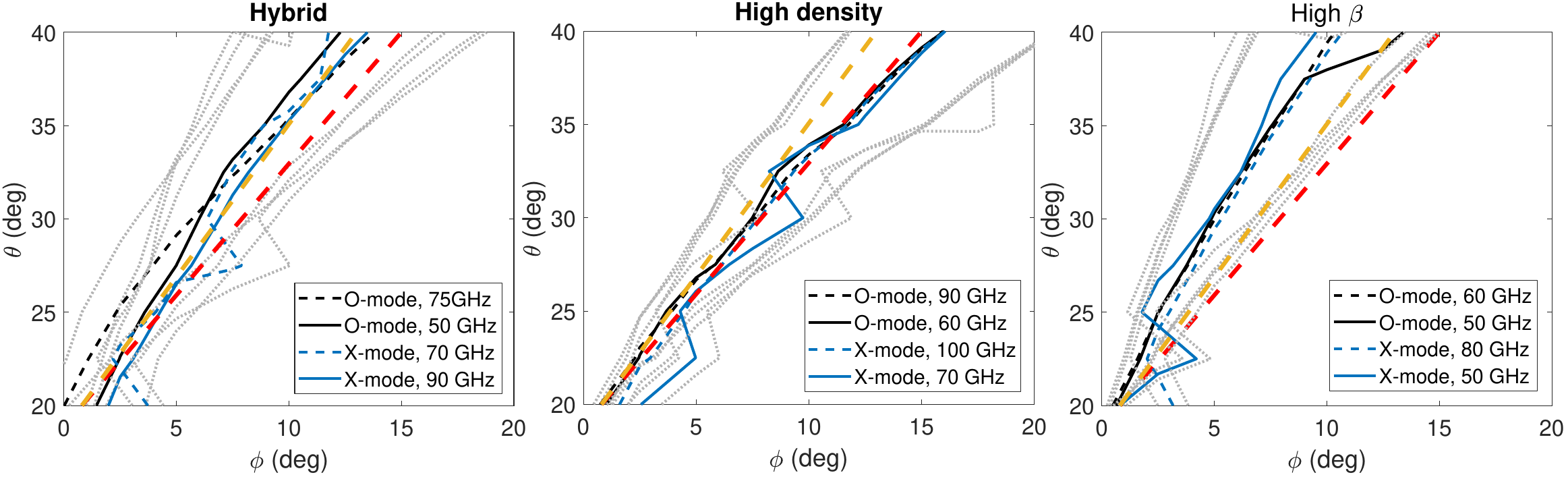}
	\caption{\textit{Optimal angle comparison. For each scenario, the optimal $\theta/\phi$ pairs (leading to k|| = 0) are represented for a number of relevant polarization/frequency combinations, as indicated in the corresponding legend. The |k$_\parallel$/k$_\perp$| < 0.1 regions are indicated as thin grey lines, equivalent to the white contours in Fig. \ref{Figure_7}. The optimal probing space for the high density configuration ($\theta = 1.41 \phi +20$) is represented as a dashed red line. The intermediate probing space ( $\theta = 1.64\phi + 18.7$) is represented as a dashed orange line.}}
	\label{Figure_9}
\end{figure}

The possibility of using a single-axis steering is a substantial advantage over a general, two-axis system due to the minimization of the complexity of the mechanical and control systems. Unfortunately, as shown in Fig. \ref{Figure_5}, the pitch angle of each configuration is  different and, as a consequence, the optimal $\theta$/$\phi$ ratio, although also roughly constant for the other scenarios, changes between them. In order to find out if a common set of angles leading to acceptable |k$_\parallel$/k$_\perp$| ratios in all three scenarios, the |k$_\parallel$/k$_\perp$| < 0.1 regions of several representative polarization and frequency combinations have been represented together in Fig. \ref{Figure_9}. As can be seen, the optimal solution for the high density scenario (represented as a dashed red line, as in Figs. \ref{Figure_7} and \ref{Figure_8}) works also reasonably well in the hybrid scenario, despite the pitch angle difference. However, it results in |k$_\parallel$/k$_\perp$| > 0.1 for almost all angles in the high $\beta$ scenario. In order to have a common set of angles for all three scenarios, an intermediate probing space with a higher pitch angle $\theta = 1.64\phi + 18.7$ is defined (displayed as an orange, dashed line in Fig. \ref{Figure_9}): as can be seen, these angles are suboptimal in the high density scenario, but get better results \textcolor{black}{than those represented by the red line} in the hybrid and meet the |k$_\parallel$/k$_\perp$| < 0.1 criterion for a wide array of angles in the high $\beta$.\\

	\begin{figure}
		\centering
		\includegraphics[width=.65\linewidth] {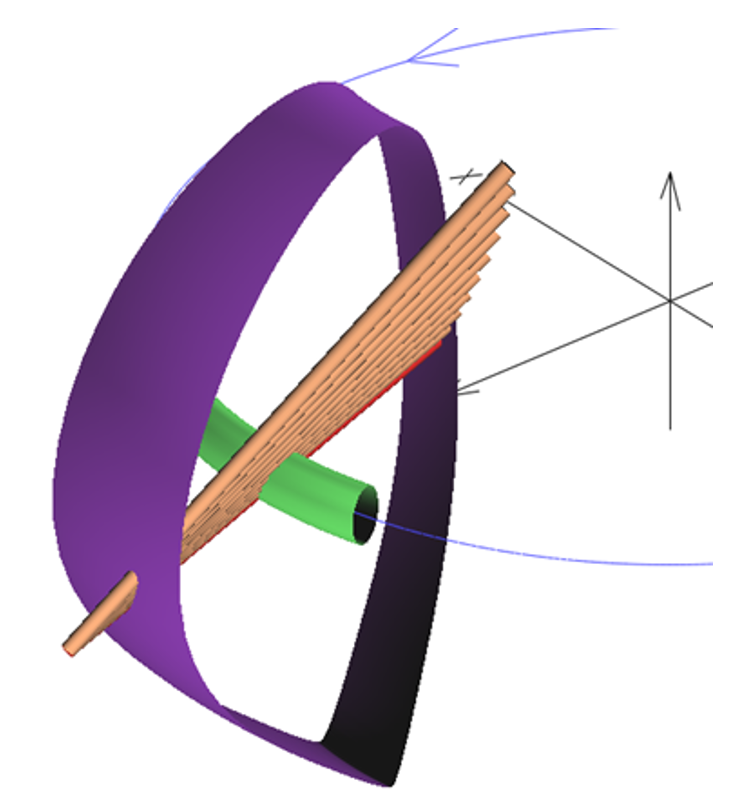}
		\caption{\textit{Second TRAVIS scan. Beams are roughly aligned over a plane featuring a constant elevation to azimuth ratio to keep the cut-off position in the optimal |k$_\parallel$/k$_\perp$| region and simulate a single-axis rotating mirror.}}
		\label{Figure_10}
\end{figure}

\subsection{Second TRAVIS scan: single-axis rotating mirror}\label{secondTravis}

This intermediate probing space is hereby used to simplify the analysis and make it closer to the design condition of a realistic DR system: a second round of scans is carried out in which a DR equipped with a single-axis rotating mirror is simulated by keeping the launching angles in the optimal region imposing the $\theta = 1.64 \phi +18.7$ condition, represented in Fig. \ref{Figure_9} as a dashed, orange line. For each of the $\theta$,$\phi$ angle combinations in the displayed array of rays, a beam with frequencies covering the whole V and W bands are launched, both with O and X-mode polarization. This way, for every polarization, frequency and scenario the ray tracing simulation is repeated for the set of launching angles delivering optimal |k$_\parallel$/k$_\perp$| ratios. \textcolor{black}{A word of caution is in order, though: while a low |k$_\parallel$/k$_\perp$| ratio is a necessary criterion to ensure good data quality, it can´t be considered a sufficient one. Therefore, a more detailed analysis would be required to ensure the feasibility of some of the measurement points. In particular, radial positions too deep into the core may suffer a degradation in spectral resolution caused by large flux surface curvatures \cite{Hirsch04} (although this problem may be less severe than in most  present machines, given the large size of JT-60SA). Similarly, for low incidence angles, $\omega_D$ is reduced along with the k$_\perp$ value and the reflected beam may reach the antenna, thus being measured along the Doppler-shifted, backscattered signal, effectively merging the two and substantially difficulting the analysis of the later. While these problems will need to be addressed in future stages of the design process, dealing with them is considered out of the scope of the present study, for which the |k$_\parallel$/k$_\perp$| ratio is expected to provide a reasonable assessment of the radial and spectral regions accesible to the diagnostic.}

\subsubsection{High Density Scenario}

\begin{figure}
	\centering
	\includegraphics[width=\linewidth] {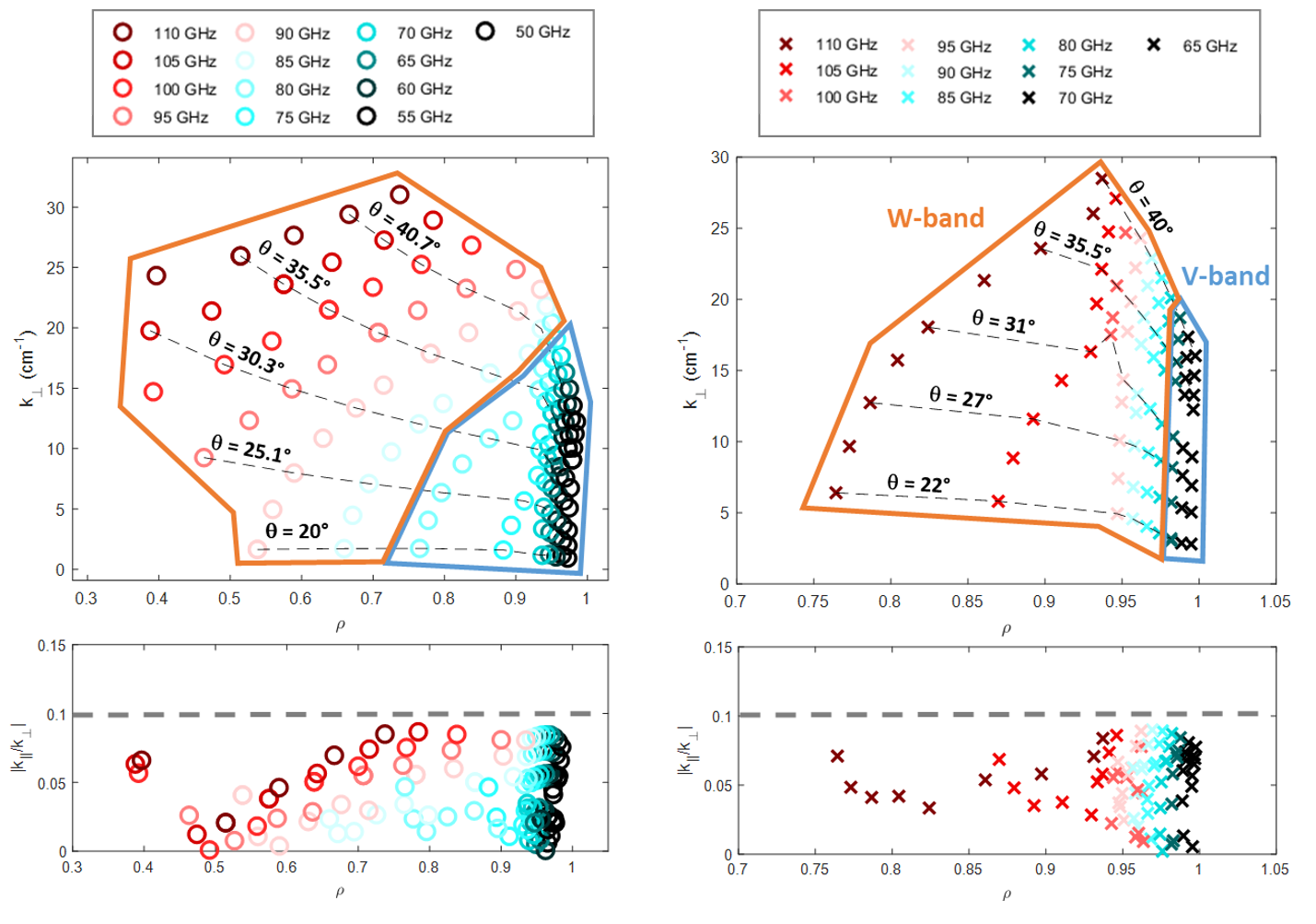}
	\caption{\textit{Results of the second scan for the high density scenario. Top: k$_\perp$ values obtained for O-mode (left) and X-mode (right) polarization. Colours stand for beam frequencies. Dashed lines connect different frequencies corresponding to the same launch angle. Microwave bands are highlighted by colored contours. Bottom: |k$_\parallel$/k$_\perp$| ratio corresponding to the points above.}}
	\label{Figure_11}
\end{figure}

Results for the high density scenario are summarized in Fig. \ref{Figure_11}, where the wave number of the turbulence and the radial position of the cut-off layer are represented for a number of beam frequencies (covering both V and W bands) and elevation angles in the O and X-mode polarizations. As well, as a measure of the quality of the signal, the parallel to perpendicular wave number ratio for each of the points are represented in separated plots. As explained above, |k$_\parallel$/k$_\perp$| < 0.1 is taken as a rule of thumb to ensure that enough S/N ratio is obtained in the measurement and therefore any points with higher ratio are excluded from the figure. 

From the results in Fig. \ref{Figure_11}, a number of conclusions can be achieved:

\begin{itemize}
\item [-] O-mode polarization provides a good radial coverage of the core, with frequencies in the W-band covering almost a decade of k$_\perp$ values in the $\rho$ = 0.4 - 0.9 range. As well, frequencies in the V-band provide a rather dense coverage of the steep pedestal region in the $\rho$ = 0.9 - 0.95 range.
\item [-] 	X-mode polarization covers a more external region, with W-band in the $\rho$ = 0.8 - 0.95 range. Instead, the V-band provides a very detailed cover of the whole pedestal, including some parts of the SOL\footnote{This is not shown in the figure, as the current version TRAVIS can´t trace rays in the SOL. However, since the frequency of the beams backscattering at the separatrix is approximately 65 GHz, it´s clear that lower V-band frequencies (down to 50 GHz) would have their cut-off at the SOL. }. 
\item [-] 	In the case of O-mode polarization, a decade-wide spectral coverage can be achieved both at the edge and core, with turbulence wave number values of k$_{\perp, core}$ = 1.5-30 cm$^{-1}$ and k$_{\perp, edge}$= 1-20 cm$^{-1}$ with |k$_\parallel$/k$_\perp$|< 0.1. As well, the |k$_\parallel$/k$_\perp$| ratio is well below the critical value in both polarizations, indicating that the alignment is particularly good for the whole range of angles considered.
\end{itemize}

\subsubsection{Hybrid scenario}

\begin{figure}
	\centering
	\includegraphics[width=\linewidth] {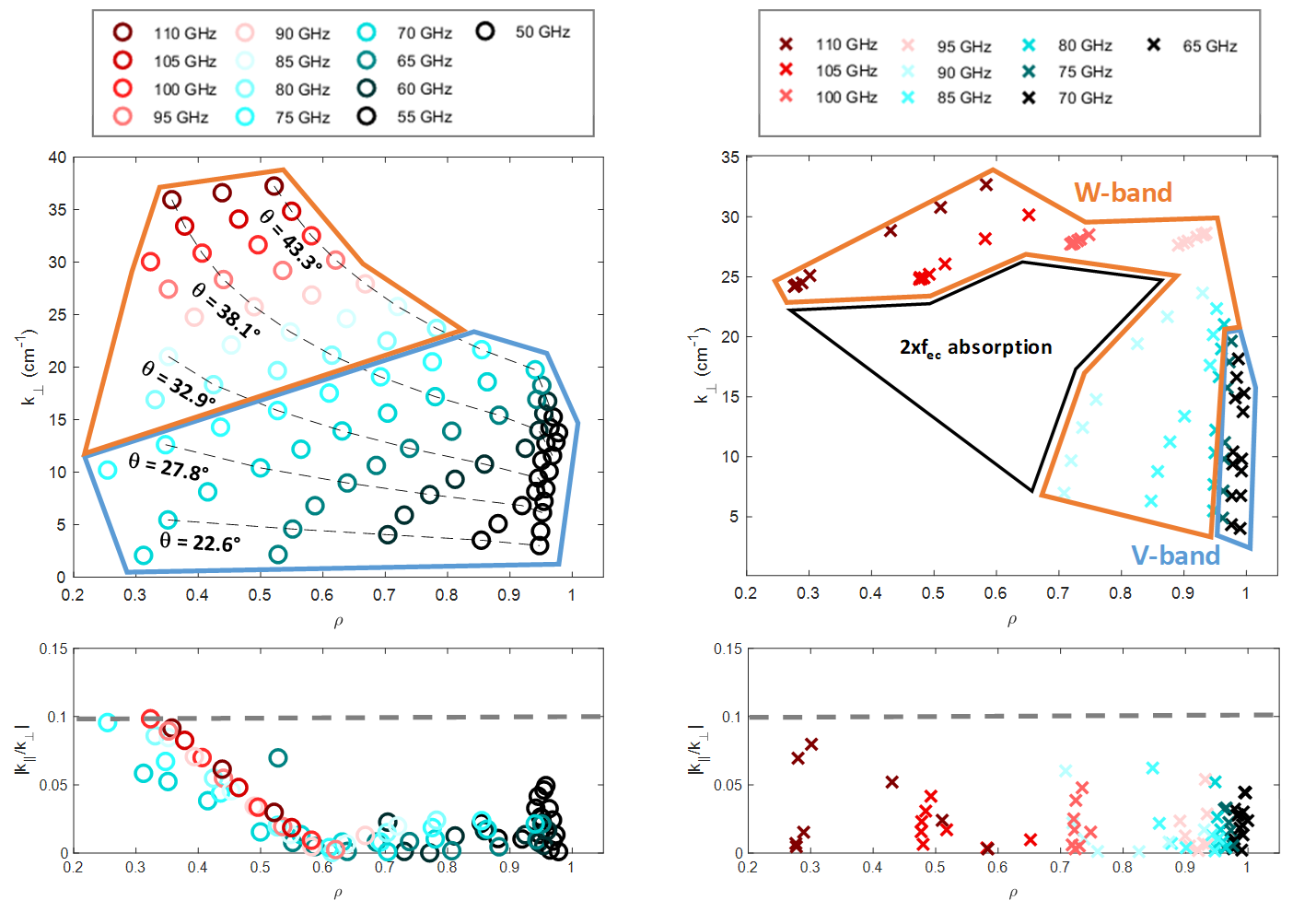}
	\caption{\textit{Results of the second scan for the hybrid scenario. Layout as in Fig. \ref{Figure_11}.}}
	\label{Figure_12}
\end{figure}

Results from the hybrid scenario are summarized in Fig. \ref{Figure_12}, with the same layout as in Fig. \ref{Figure_11}. In this case, the lower densities allow deeper cut-off layers for the same range of frequencies, although the |k$_\parallel$/k$_\perp$| ratio increases for the innermost radii. As well, a large empty region appears in the X-mode polarization case, as the beam crosses the second harmonic of the electron-cyclotron resonant region, in which the beam is absorbed and no signal is scattered back. As a result, most of the W-band frequencies are useless and the core can´t be probed with this polarization mode. The main results from this second scan are then:
\begin{itemize}
\item [-]	The central region can be completely covered by the V-band in O-mode. However, in order to get a good spectral coverage, a few extra frequencies are required (ideally 75-90 GHz, making the E-band the best suited for the observation of turbulence in this scenario).
\item [-]	The edge would be covered by both O-mode and X-mode polarizations in V-band with a reasonably good spectral range. 
\item [-]	While X-mode beams can´t reach most of the core region due to absorption, they would cover the edge adequately and provide additionally an extended coverage of the SOL. 
\item [-] As in the high density scenario, by optimizing angles and frequencies, wide spectral coverage can be achieved, while low |k$_\parallel$/k$_\perp$| ratios are kept for most of the displayed radial range.
\end{itemize}

\subsubsection{High beta scenario}

\begin{figure}
	\centering
	\includegraphics[width=\linewidth] {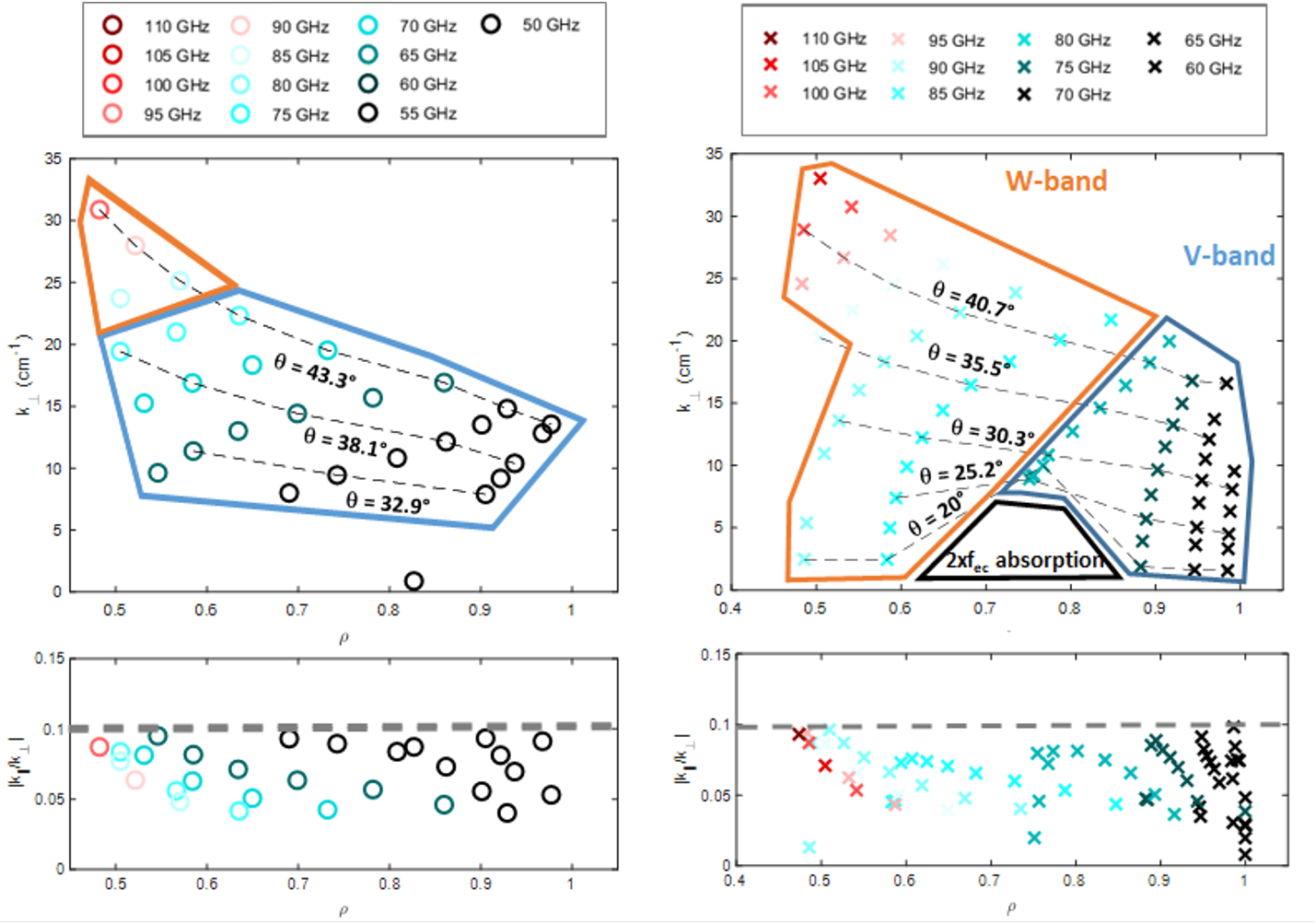}
	\caption{\textit{Results of the second scan for the high $\beta$ scenario. Layout as in Fig. \ref{Figure_11}.}}
	\label{Figure_13}
\end{figure}

Results from the high $\beta$ scenario are summarized in Fig. \ref{Figure_13}, also with the same layout as in Fig. \ref{Figure_11}. This scenario features a density profile similar to that of the hybrid scenario, resulting in similar behaviour in the O-mode polarization case. However, as seen in Fig. \ref{Figure_9}, the selected angles are not as optimized for the pitch angle profile of this scenario, resulting in a more limited set of angle pairs meeting the |k$_\parallel$/k$_\perp$|  < 0.1 condition. In contrast, the valid region for the X-mode polarization is slightly wider meaning that most of the angle pairs are within such condition and yielding a good coverage of the profile both in the radial and spectral range. The second harmonic absorption region is also present in this scenario, but its influence is substantially less deleterious than in the hybrid, being limited to a handful of frequencies for low values of $\theta$. The main conclusions from this scan would be:
\begin{itemize}

\item [-]	 O-mode polarization in V-band covers the whole radial profile from the LCFS to $\rho$ = 0.5. However, given that the defined $\phi$,$\theta$ set of angles is suboptimal for this scenario (see Fig. \ref{Figure_9}) the range of accessible k$_\perp$ values is considerably limited with respect to other scenarios. 
\item [-]	X-mode polarization allows for a reasonably good spectral coverage of the core and edge in W-band and V-band, respectively, with k$_\perp$ = 2-30 cm$^{-1}$ for $\rho$ = 0.5-0.6 and k$_\perp$ = 2-20 cm$^{-1}$ for $\rho$ = 0.9-1.  
\item [-]	Also as a result of the sub-optimal selection of angles, |k$_\parallel$/k$_\perp$| ratios tend to be higher than in the other scenarios. Fortunately, a reasonably wide range of measurements can be achieved with this intermediate solution. 
\end{itemize}

\subsubsection{Discussion}\label{disc_3}
 \textcolor{black}{The main results from the previous subsections have been summarized in Fig. \ref{Figure_4}, where the radial ranges accessible to the DR in each scenario have been represented for the k$_\perp \simeq 10$ cm$^{-1}$ case}. The main conclusion is that a common geometry for the antenna/mirror system exists such that, with a single axis of rotation, both core and edge can be probed for a roughly decade-wide range of turbulence wavenumber values and low |k$_\parallel$/k$_\perp$| values for all three scenarios. This holds true despite the considerable variation of both magnetic configurations (pitch angle, |B| values) and plasma density profiles between them. A follow-up conclusion is that a very versatile system could be designed with just a single-axis rotating mirror and two reflectometers: 
\begin{itemize}
\item [-] The first one, using the V-band of frequencies and polarized in the X-mode, would be mainly designed for the observation of the edge and SOL, which is adequately covered by this combination in all three scenarios. The fact that some region of the SOL can be covered -even in scenarios featuring lower densities- is particularly advantageous since that should allow the measurement of the large E$_r$ shear typically found around the separatrix (which not only is a very relevant physical quantity in itself, but also provides a good reference point for the radial localization of the reflectometer profile). As well, since this channel would be dedicated mainly to the observation of turbulence and rotation around the ETB, X-mode polarization would be preferred as it provides better spatial localization in this step gradient region.

\item [-]	The second one, polarized in the O-mode would be mainly used for the observation of the core, and for the characterization of turbulence in it. In this case, it is also advantageous to use O-mode polarization in order to reduce the levels of non-linear wave-plasma interactions \cite{ref13,ref33} and specially to avoid the absorption region associated to the second harmonic of the X-mode appearing in some scenarios at the outer region of the core. In this case, the optimal frequency range can’t be defined so easily: in the high density scenario, the W-band would be required, whereas in the others the lower density profiles reduce the required frequencies, with the optimal band being somewhere between the E and the W.
\end{itemize}

\begin{figure}
	\centering
	\includegraphics[width=.75\linewidth] {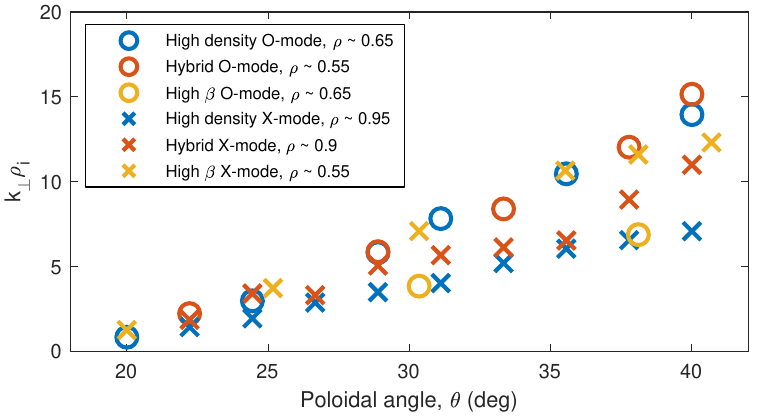}
	\caption{\textit{Wave numbers of the back-scattering turbulence are normalized using the Deuterium gyroradius at the cut-off layer for a number of cases in the elevation angle range discussed in Figs. \ref{Figure_11}-\ref{Figure_13}. Symbols/Colours stand for polarization modes/scenario, as described in the legend. }}
	\label{Figure_14}
\end{figure}

The design of the second reflectometer may thus require a decision regarding which scenarios are considered the most relevant for the characterization of turbulence, introducing a compromise solution which would favour a certain range of radial positions and density ranges at the expense of others. Perhaps a better solution would be to design the system in such a way that the polarization of the two reflectometers could be exchanged. In that case, one reflectometer would use the V-band and the other the W-band, and their respective polarizations would be adjusted depending on the experimental requirements, thereby providing a rather complete cover of the whole plasma in all scenarios. A second factor to consider in such more detailed design is the measurement of rotation at the core. However, it can be readily seen from the scans that, if a proper frequency band is selected for obtaining a wide range of k$_\perp$ values at the core, the maximum radial range of measurements is also achieved. Finally, the scientific relevance of the observed wavenumbers is assessed by discussing the k$_\perp\rho_i$ values of the back-scattering turbulence, displayed in Fig. \ref{Figure_14}. As can be seen, the k$_\perp\rho_i \simeq$ 0.5-15 range is covered at the core for all scenarios (in the high $\beta$ one, X-mode is used for this). \textcolor{black}{These values are rather close to the ones proposed initially, and should be sufficient to measure fluctuations corresponding to ion-scale as well as TEM turbulence. Nevertheless, two additional conclusions can be obtained from the analysis:
	\begin{itemize}
		\item [-] In the first place, the obtained values of k$_\perp\rho_i$ are still slightly over the ones for which maximum amplitude is expected for ITG turbulence k$_\perp\rho_i < 0.5$. This is particularly the case at the edge or in the hybrid scenario, for which the minimum value is closer to k$_\perp\rho_i \simeq 1$. In order to extend the measurement range in this direction, more perpendicular incidence angles would need to be explored, which would require a more detailed analysis, considering if the resulting backscattered Doppler peak can still be separated from the reflected beam. It must be taken into account that, as can be seen in Fig. \ref{Figure_9}, if $\theta$ and $\phi$ are reduced, the range in which the k$_\parallel$/k$_\perp < 0.1$ condition holds is considerably reduced and begin to diverge for different configurations. As a result, such more detailed analysis would also be required to ensure that lower k$_\perp$ values can be achieved with sufficient signal quality. Most likely, this will only be possible for a given scenario, which will have to be selected based on its scientific interest. Both tasks are left for future work. \\
		\item[-] Secondly, from the perspective of the geometrical analysis, rather high values of k$_\perp\rho_i$ up to 15 seem to be accessible for reasonable incidences. While k$_\perp\rho_i > 10$ have been reported in the literature \cite{Hillesheim15}, it remains unclear if turbulence amplitude in JT-60SA will be high enough to obtain useful measurements at this scales. Should that be the case, it would open the door to the exploration of electron-scale turbulence. While this was not initially considered among the scientific objectives of the project, it should be considered as an interesting opportunity to widen its scope.
		\end{itemize}
}

 \textcolor{black}{
As a closing remark, it should be noted again that, while the k$_\parallel$/k$_\perp$ criterion followed in this analysis has allowed for a first order discussion of the feasibility of a DR system for JT-60SA, a number of important aspects have been left out of its scope and will need to be addressed in future, more detailed iterations. Some of these have already been mentioned, such as the effect of flux surface curvature on spectral resolution for the innermost radii, the separation of backscattered and reflected beams for low k$_\perp$ values and the need for careful angle selection in order to properly cover the ITG turbulence scale. Without being exhaustive, other aspects to be considered in the future would include an evaluation of scattering efficiency (which may be a problem for high k$_\perp$ values), the paramatric range in which wave-plasma interactions can be expected to transit into the non-linear regime \cite{Gusakov02} or, given the high T$_e$ values present at the accesible radial range, the importance of relativistic corrections.
}

\section{Preliminary conceptual design proposal}\label{Sec3}

In the previous section, an optimal geometrical design has been determined for the given launching position. In this one, we will discuss preliminarily how such ideal design could be translated into a real conceptual design of a working diagnostic. However, since no port has been definitively allocated for the installation of a DR system at the time of this work, this second phase is conditioned by some uncertainties:
\begin{itemize}
\item [-]	Launching position for the DR.
\item [-]	Volume available for the diagnostic components outside the vacuum vessel.
\item [-]	Volume available for the diagnostic components inside the vacuum vessel
\item [-]	Mechanical/volumetric constrains associated to the shape of the port plug used to install the diagnostic.
\end{itemize}

Regarding the first, we will simply continue to assume the same position as in the previous section. This doesn´t seem to be a critical issue, though: as long as the DR can be installed in some port roughly placed around the outer midplane, it seems likely that a similarly optimal solution could be obtained by repeating the previous calculations. Regarding the second, it is assumed that sufficient space is allocated in the vicinity of the port (such as the one displayed Fig. \ref{Figure_2}). Therefore, waveguides between the wave generators and the ports are assumed to be a few meters long, with the subsequent savings in signal attenuation and material costs.\\

Unfortunately, no such simple hypothesis can be made regarding the third and fourth constrains, as they depend critically on the details of the final port space allocated for the diagnostic. In this situation, the following approach has been decided: first, a worst-case scenario has been assumed, in which only one of the lower circular ports are made available. Starting from this premise, a minimum viable system is proposed and discussed. Then, taking such design as a starting point, possible improvements are discussed along with the space requirements for each variant of the system.\\

\subsection{Minimum Viable System}
The minimum viable system is designed starting from the following premises:
\begin{itemize}
\item [-]	The whole system is installed inside the bottom left circular port displayed in the left plot of Fig. \ref{Figure_2}. 
\item [-] 	The inner diameter of the port plug is 200 mm. The available length of the port plug is at least 2 m.
\item [-] 	The whole system can be extracted with the port plug.
\end{itemize}

\begin{figure}
	\centering
	\includegraphics[width=0.75\linewidth] {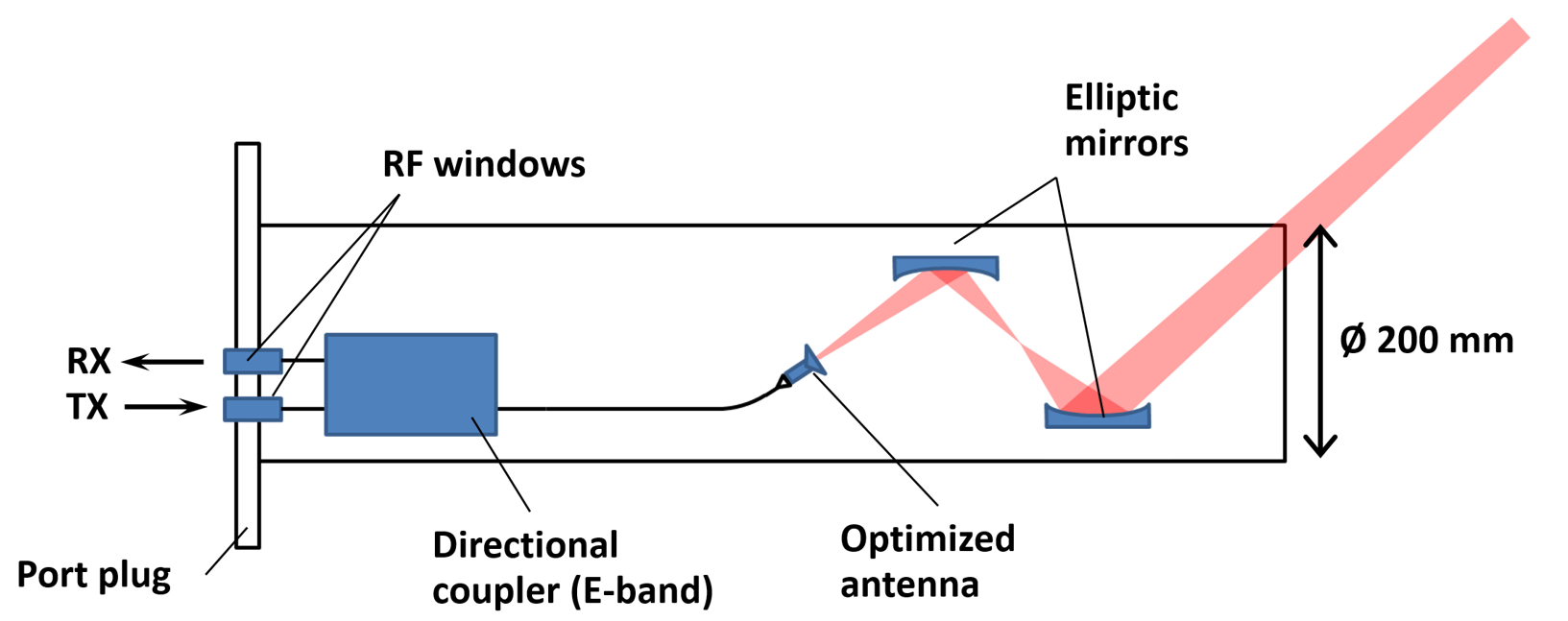}
	\caption{\textit{Overview of a minimum viable DR system to be installed in the lower left circular port plug of the P-18 Horizontal diagnostic flange, as seen in Fig. \ref{Figure_2}. }}
	\label{Figure_16}
\end{figure}

Taking into account these conditions, a DR system could be designed following the schematic in Fig. \ref{Figure_16} \textcolor{black}{, representing a port plug to be inserted in the aforementioned port of Fig. \ref{Figure_2}, with the position of the last elliptic mirror corresponding to the launching point indicated as a red cross in Fig. \ref{Figure_3}}: From left to right, a couple of RF windows connect the outer waveguides to the vacuum thigh part of the port plug. These are then connected by fundamental V-band waveguides to a E-band directional coupler. In the third port of the coupler, another fundamental V-band waveguide is connected by a rectangular to circular taper to a circular antenna at the other end. All these components have been designed in order allow the use of V-band, E-band or W-band reflectometers without severe signal degradation, thereby keeping in principle the frequency range displayed in Fig. \ref{Figure_11} to \ref{Figure_13}. \textcolor{black}{The selection of an E-band coupler comes from the necessity to accomodate two different bands (V and W) with a single device. This is obviusly suboptimal and may lead to moderately lower S/N ratios outside the nominal working frequencies of the coupler.}  The purpose of the fundamental V-band waveguide after the coupler is to allow for a smooth bend of its axis which provides the right orientation for the antenna without generating spurious modes (as would be the case with an oversized waveguide).  The antenna would feature a choked-corrugated design, which would allow for V-E-W band operation within the aforementioned space restrictions. \\

Regarding the optical design, a monostatic system has been selected to reduce the number of components thus allowing for bigger mirrors. This is required in order to reduce spillover and achieve a larger beam waist at the cut-off layer. A preliminary design of the system includes two elliptic mirrors of ca. 100 mm of diameter and f = 100 mm and f = 400 mm. They are placed 50 mm above and below the axis of the port plug and with a longitudinal distance of 500 mm between them, forming a Gaussian telescope in order to obtain a frequency-independent setup. Preliminary calculations carried out for a Gaussian beam with an intermediate frequency (90 GHz) and a 7.5 mm waist (narrow enough to be launched from the proposed antenna) yield a final 30 mm waist width at the measurement point (around 1.4 m away from the last mirror). Finally, regarding the components outside the vacuum vessel, two reflectometers (ideally, one operating in the V-band and the second in the W-band) could be installed in a conventionally sized rack and connected by waveguides to the port plug. There, both microwave beams would be combined in a beam splitter and passed through the RF windows. The same arrangement would be set for the RX part of the circuit. \\

This setting has its main advantage in being completely functional, allowing the simultaneous use of both V and W bands yet being rather compact and thus fitting in the reduced available space. On top of that, the Gaussian telescope arrangement implies that the final beam waist would not change due to the quasioptical transmission, its only dependence with frequency being that of the radiation pattern of the antenna. Of course, the reduced space still imposes a number of limitations, the main ones being:
\begin{itemize}

\item [-] 	Directional couplers have a fixed polarization, meaning that either O or X-mode polarization would need to be fixed by design, and remain unchanged unless the port plug can be extracted. This would substantially limit observations at the core or the edge, depending on the selected solution. 
\item [-]	There is not enough space inside the port plug to allow for antenna/mirror steering. As a result, the injection angle would also be fixed by design and no wave number spectrum measurements would be possible. 
\item [-]	Relatively shallow incidence angles in the mirrors may cause beam to depart from gaussianity. More detailed optimization of the mirrors and their layout would be required to prevent this.

\end{itemize}

 \textcolor{black}{In summary, this minimum viable system would allow for DR measurements to be carried out, but at the price of restricting its operation to a certain set of scenarios for a given campaign, and imposing very severe limitations at the scope of the achievable scientific program.}

\subsection{Improved Designs}

The minimum viable system would allow for fast measurements of ion-scale (ITG,TEM) turbulence amplitude and radial electric field profiles, in a \textcolor{black}{limited range of positions either at the core or the edge}. From it, a number of additional features could be included if more space was available and/or if the geometric restrictions imposed from the port were at least partially lifted. Ordered by their scientific importance, these improvements would be:
\begin{itemize}

\item [-]	Allow sufficient space for an additional directional coupler (with its corresponding two RF windows at the flange), allowing the use of the two polarization modes. This design would also allow for a relatively easy combination of the two frequency bands and O/X polarization, as exchanging the bands could easily be done remotely in a stage prior to the RF windows. \textcolor{black}{This improvement would greatly enhance the radial range and operational flexibility of the diagnostic, essentially allowing it to cover the radial regions described for each scenario in Fig. \ref{Figure_4}}.
\item [-]	Installing a steerable mirror to control the launching angle. This improvement would allow the kind of k$_\perp$ spectra measurements discussed in the previous section and would be needed in order to carry out any detailed comparison with non-linear GK simulations. It would require a moderate increase in the available space, as the last mirror would now be allowed to rotate along one axis (and this might require some minor repositioning of the rest of the components). As well, a mechanical actuator would be installed at the outer side of the flange and connected to the mirror by a rod and a feedthrough in order to change its position remotely.
\item [-]	Installing an additional channel in the existing port plug would allow for the measurement of radial correlations and a further detailed characterization of turbulence, including the tilt angle of eddies which would allow for more meaningful comparison to GK codes. This would require a third reflectometer system to be installed in the outer rack, as well as some minor microwave components.  
\item [-]	Finally, installing a separate antenna in a second toroidally/poloidally separated port. This improvement would allow for the measurement of long range correlations for the detection and characterizations of ZF. It would require a system equivalent to the minimum viable one installed in a similarly sized port plug with a different toroidal and/or poloidal position. If more space could be allocated for this system, and it would be equipped with a steerable mirror, it would also allow for the evaluations of poloidal asymmetries in the turbulence. 

\end{itemize}

\subsection{Space Requirement Assessment}

In order to carry out an assessment of space requirements of the system, four scenarios have been considered:
\begin{itemize}

\item [-]	Bare minimum scenario, in which only the minimum viable system is installed.
\item [-]	Baseline scenario, in which some additional space is made available for the DR system in the P-18 Horizontal port allowing for the installation of a second directional coupler and a steering system.
\item [-]	Extended scenario A, in which a third reflectometer for radial correlations is added to the Baseline scenario.
\item [-]	Extended scenario B, in which a second port is designated for the installation of an additional DR system for the measurement of poloidal/toroidal correlations.

\end{itemize}

The bare minimum scenario assumptions have been already discussed in the subsection dealing with the minimum viable system. The baseline scenario assumes that the lighting system currently installed in the two lower ports of the flange is removed, thereby liberating a surface of ca. 700x400 mm. In such case, different designs could be developed, depending on the level of modification allowed on the flange: if the two ports are made available, but their plugs must be kept as they are now, the minimum viable system could be replicated using a coupler with different polarization \textcolor{black}{(eg., one focused on the edge and the other on the core, as discussed in section \ref{disc_3})}, but there would be probably not enough space for a steering system. On the other hand, if the whole space was turned into a large new port, there would more than enough space for the baseline scenario design, and room for additional diagnostics could be probably included in the design. A good candidate for this would be a different reflectometry system featuring a lower band of frequencies for the observation of outer edge and SOL physics. Such a system has already undergone a feasibility study \cite{Tokihito} and would fittingly complement the current proposal, providing a full SOL-to-core coverage of E$_r$ and fluctuation amplitude measurements. Extended scenario A would probably require just a minor increment in the used space in the torus hall. Depending on the size of the rack, it might be installed without further expansions. Extended scenario B would require a second port plug to be dedicated to the DR system, with a size at least equivalent to the one dedicated to the minimum viable system.\\

\section{Summary of conclusions}\label{Sec4}

The main conclusions of the present study are the following:

\textbf{1. A Doppler reflectometer would be a very relevant diagnostic for the accomplishment of the JT-60SA scientific program.} The JT-60SA tokamak has been designed as part of the Broader Approach agreement to fusion power, with the main objective of reproducing relevant scenarios from which the operation of ITER can be anticipated and the demonstration of integrated performance scenarios for ITER and DEMO. With this aim, JT-60SA will achieve plasmas with features much closer to ITER and DEMO than those of present-day machines. However, the projection of observations carried out in JT-60SA onto ITER and DEMO-relevant scenarios will require the validation of current models of fundamental plasma phenomena. One of the most prominent examples of this is the evaluation of turbulence models under new plasma parameters ($\beta$, $\nu^*$, $\rho^*$, etc..), as well as related phenomena such as transport barriers, L-H transition, etc. However, other phenomena (such as rotation and neoclassical toroidal viscosity) would require similar model verification in order to confidently predict scenarios in reactor-relevant settings. A DR system is one of the best options available for this mission, as it can carry out the kind of well-localized, quantitative characterization of turbulence required for the comparison with GK codes. Also, this diagnostic provides simultaneous measurements of the stabilizing role of E$_r$ and E$_r$ shear in turbulence, particularly relevant for some phenomena such as the ETG associated to the pedestal formation in the L-H transition. Finally, it can provide toroidal flow profiles independently of NBI heating, which may be of paramount importance for the validation of neoclassical toroidal viscosity (NTV) models under high beta scenarios in which rotation will be critical for the stabilization of MHD modes.\\

\textbf{2. It is possible to build a Doppler Reflectometer in JT-60SA capable of achieving its scientific objectives.} The analysis presented in this report has identified a geometric solution for the installation of a Doppler Reflectometer such that both edge and core can be observed and wave number spectra can be measured using a single-axis steering system, while keeping reasonably low |k$_\parallel$/k$_\perp$| values (indicating good S/N ratios in the measurement). In particular, such a system would consist of two independent reflectometers that could measure simultaneously:
\begin{itemize}
\item [-]	A W-band channel would cover the core under most of the analysed scenarios (although frequencies in the E-band are also required under the ones featuring lower densities). 
\item [-]	A V-band channel would cover with great detail the pedestal region in all analysed scenarios. 

\end{itemize}

The flexibility of such system would be greatly enhanced if each of these bands can be polarized either in O or X-mode: First, this would allow for density fluctuation amplitude and radial electric field measurements from the inner core ($\rho \simeq$ 0.35) to the SOL, as well as turbulence spectra observation in at least the k$_\perp\rho_i\simeq$ 0.5-15 range at the core and k$_\perp\rho_i\simeq$ 1-10 range at the edge under the most relevant operation scenarios. Second, operation would benefit from the special characteristics of each polarization depending on the quantity of interest for a given experiment: O-mode features lower non-linear plasma-wave interaction, thus reducing the error when measuring turbulence amplitude. Instead, X-mode polarization features better spatial localization, which can be useful, eg., when measuring rotation in the steep gradients of the pedestal.\\ 

\textbf{3. A conceptual design could be carried out if a fraction of a Horizontal port is allocated for this diagnostic.} In this report, several possibilities for a conceptual design have been discussed. In the first place, a minimum viable system has been proposed, in which important scientific features have been traded off in order to achieve a sufficiently compact design, which could fit in one of the two lower circular port plugs of the P-18 Horizontal port. However, due to limitations of space, this design would lack the capability to steer its last mirror, thereby giving up the capability to measure fluctuation spectra. The best return for investment would be achieved with a moderately larger system, which would retain all the features proposed initially. This baseline design would require a fraction of the ca. 700x400 mm region currently dedicated for the illumination system in the P-18 horizontal port. Ideally, if the current two 300 mm diameter ports are replaced by a larger port, the resulting available space would exceed that required for the baseline design, which might be then made compatible with additional systems/diagnostics. As well, if this baseline design is selected, an additional reflectometer could be included with no extra space requirements in the vessel to carry out correlation studies to improve turbulence characterization (Extended design A). Finally, if an additional port is made available with space in it at least equivalent to the one for which the minimum viable system has been designed, measurement of long range correlations for the detection and characterizations of ZF would be possible (Extended design B).\\


\section*{Acknowledgments}

The authors gratefully acknowledge members of the JT-60SA Integrated Project Team for data exchange and fruitful discussions. This work has been partially funded by the Spanish Ministry of Science and Innovation under contract number FIS2017-88892-P and sponsored in part by the Comunidad de Madrid under project 2017-T1/AMB-5625. This work has been carried out within the framework of the EUROfusion Consortium and has received funding from the Euratom research and training programme 2014-2018 and 2019-2020 under grant agreement No 633053. The views and opinions expressed herein do not necessarily reflect those of the European Commission.\\

\end{document}